\documentclass[preprint,5p,times,twocolumn]{elsarticle}

\pdfoutput=1 

\usepackage{mathrsfs}
\usepackage{amsmath}
\usepackage{amsfonts}
\usepackage{color}
\usepackage{graphicx}
\usepackage{graphics}
\usepackage{amssymb}

\journal{Elsevier}

\begin{document}

\begin{frontmatter}



\title{Competing s-wave orders from Einstein-Gauss-Bonnet gravity}

\author[label1]{Zhi-Hong Li,}
\author[label1]{Yun-Chang Fu,}
\author[label1,label2]{Zhang-Yu Nie\corref{cor1}}
\ead{niezy@kmust.edu.cn}

\cortext[cor1]{Corresponding author}



\address[label1]{
Kunming University of Science and Technology,
\\No. 727 South Jingming Road, Cheng Gong District, Kunming, 650500 China}
\address[label2]{Key Laboratory of Theoretical Physics, Institute of Theoretical Physics,\\Chinese Academy of Sciences,
 P.O. Box 2735, Beijing, 100190 China}

\begin{abstract}
In this paper, the holographic superconductor model with two s-wave orders from 4+1 dimensional Einstein-Gauss-Bonnet gravity is explored in the probe limit. At different values of the Gauss-Bonnet coefficient $\alpha$, we study the influence of tuning the mass and charge parameters of the bulk scalar field on the free energy curve of condensed solution with signal s-wave order, and compare the difference  of tuning the two different parameters while the changes of the critical temperature are the same. Based on the above results, it is indicated that the two free energy curves of different s-wave orders can have one or two intersection points, where two typical phase transition behaviors of the s+s coexistent phase, including the reentrant phase transition near the Chern-Simons limit $\alpha=0.25$, can be found. We also give an explanation to the nontrivial behavior of the $T_c-\alpha$ curves near the Chern-Simons limit, which might be heuristic to understand the origin of the reentrant behavior near the Chern-Simons limit.
\end{abstract}


\end{frontmatter}








\section{Introduction}
\label{sec:intro}
The Anti-de Sitter/Conformal Field Theory (AdS/CFT) correspondence~\cite{Maldacena:1997re,Gubser:1998bc,Witten:1998qj} is believed to be a useful approach to study strongly coupled systems through weakly coupled gravitational duals~\cite{Sakai:2004cn,Karch:2006pv,Zaanen:2015oix,Cai:2015cya}. One of the most successful application of this duality is the so-called holographic superconductor, which was first realized in refs.~\cite{Gubser:2008px,Hartnoll:2008vx} to mimic an s-wave superconductor with a charged scalar field in bulk.

Recently, the technique of realizing various phase transition order parameters in holographic superconductor models have been applied to the study the competition and coexistence of multiple order parameters~\cite{Basu:2010fa,Cai:2013wma,Nie:2013sda,Nie:2014qma,Musso:2013rnr,Chaturvedi:2014dga,Huang:2011ac,Donos:2011ut,Krikun:2012yj,Donos:2012yu,Nitti:2013xaa,Liu:2013yaa,Wen:2013ufa,Amado:2013lia,Amoretti:2013oia,Momeni:2013bca,Donos:2013woa,Nishida:2014lta,Li:2014wca,Kiritsis:2015hoa,Wu:2015sqk,Chen:2016cym,Li:2016uhr,Arias:2016nww,Li:2017wvw}. The authors of ref.~\cite{Basu:2010fa} studied the competition effect of two scalar order parameters in the probe limit and get an interesting phase with coexistent orders. In ref.~\cite{Cai:2013wma}, the authors extended the above study to cases including full back reaction of matter fields on the metric, and showed a rich phase structure. The authors of refs.~\cite{Nie:2013sda,Nie:2014qma} studied the competition between an s-wave order and a p-wave order holographically in and away from probe limit, respectively. They found that in the case away from the probe limit, more interesting phase transitions such as the reentrant one can be found for the coexistent phase.

Some of the interesting phase transitions in the system with multiple orders are useful to study universal laws in the boundary field theory. In some of these studies that involving the non-equilibrium physics~\cite{Bhaseen:2012gg,Adams:2012pj,Du:2014lwa,Du:2015zcb} or inhomogeneous effects~\cite{Liu:2015zca,Wang:2016jov}, we will encounter complex numerical work. If possible, it would be much simpler to do these studies in probe limit. Therefore it is valuable to find interesting phase transitions holographically in probe limit.

It is wise to try these studies in Gauss-Bonnet gravity, because some previous holographic studies in Gauss-Bonnet gravity show that the higher curvature terms have nontrivial contributions to some universal properties in Einstein gravity, such as the shear viscosity bound~\cite{Brigante:2007nu,Brigante:2008gz,Cai:2008ph} and the frequency gap in optical conductivity for holographic superconductors~\cite{Gregory:2009fj,Pan:2009xa,Cai:2010cv,Lu:2016smd}. Interesting multiple reentrant phase transitions and ``$\lambda$-line'' phase transition have also been realized recently in more general Lovelock gravity~\cite{Frassino:2014pha,Hennigar:2016xwd}. If the holographic models with multiple order parameters are studied in Einstein-Gauss-Bonnet gravity, it would be possible to find phase transition behaviors that are difficult to be realized in Einstein gravity in the probe limit.


In this paper, we study the competition of two s-wave orders holographically from Einstein-Gauss-Bonnet gravity in probe limit, and try to find interesting phase transition behaviors for the s+s coexistent phase. Previous studies indicate that coexistent phases can be easily found near the intersection point of the free energy curves of phases with single order~\cite{Nie:2014qma}. Therefore, we study the free energy curves of the different s-wave solutions at first. With these information, it is more easy to search the s+s coexistent solutions. Our study will also check this relation between the free energy curves of phases with single order and the emergence of coexistent phase with multiple orders.

This article is arranged as follows. In section \ref{sec:setup}, the holographic model with two s-wave orders is set up on the Einstein-Gauss-Bonnet black brane background and some details for calculation are introduced. In section \ref{sec:S1}, we study and compare the influences of the two model parameters $m_1$ and $e_1$, which are the mass and charge of the scalar filed in bulk, on the free energy curves of the phase transitions triggered by single s-wave order at different values of the Gauss-Bonnet coefficient $\alpha$. In section \ref{sec:s+s}, we show the condensate and free energy of two typical phase transition behaviors of the $s+s$ coexisting phase including the reentrant one near the Chern-Simons limit. In section \ref{sec:alpha0.25}, we explain the special behavior of $T_c-\alpha$ curve near the Chern-Simons limit which might be heuristic to understand the reentrant behavior. We conclude in section \ref{sec:Conclusions} with our main results and further discussions.


\section{The holographic s+s model in Einstein-Gauss-Bonnet gravity}
\label{sec:setup}

\subsection{Gravitational background}
We consider a holographic model with two s-wave orders from the Einstein-Gauss-Bonnet theory in (4+1) dimensional space time in the probe limit. The gravitational back ground can be taken as the 4+1 dimensional asymptotic AdS black brane solution~\cite{Cai:2001dz}
\begin{equation}\label{metric}
\text{d}s^{2}=-f(r)\text{d}t^{2}+\frac{\text{d}r^{2}}{f(r)}+\frac{r^{2}}{L^{2}}(\text{d}x^{2}+\text{d}y^{2}+\text{d}z^{2}) ,
\end{equation}
where
\begin{equation}
f(r)=\frac{r^{2}}{2\alpha}[1-\sqrt{1+\frac{4\alpha}{L^{2}}(\frac{r_{h}^{4}}{r^{4}}-1)}] ,
\end{equation}
and the position of the horizon $r_{h}$ is defined by $f(r_{h}=0)$. L is the AdS radius, $\alpha$ is the Gauss-Bonnet coefficient with dimension $(length)^{2}$. In the asymptotic region$(r\rightarrow\infty)$, the function $f(r)$ approaches
\begin{equation}
f(r)\sim\frac{r^{2}}{2\alpha}[1-\sqrt{1-\frac{4\alpha}{L^{2}}}] .
\end{equation}
Therefore, the effective AdS radius can be defined by
\begin{equation}\label{Leff2}
L_{\text{eff}}^{2}=\frac{2\alpha}{1-\sqrt{1-\frac{4\alpha}{L^{2}}}} .
\end{equation}
Note that $\alpha\leq{L^{2}}/{4}$ is known as the Chern-Simons limit.

The Hawking temperature of the black brane is given by
\begin{equation}\label{temperature}
T=\frac{r_{h}}{\pi L^{2}},
\end{equation}
which is also interpreted as the temperature of dual field theory.

\subsection{Equations of motion}
On the asymptotical AdS background, we need a Maxwell field and two charged complex scalars to study the competition and coexistence between two s-wave orders. The action of the matter fields is
\begin{eqnarray}
S_{M}&=&\int \text{d}^{5}x\sqrt{-g}\Big(-\frac{1}{4}F_{\mu\nu}F^{\mu\nu} \\ \nonumber
&&-|\nabla_{\mu}\psi_{1}-ie_{1}A_{\mu}\psi_{1}|^{2}-m_{1}^{2}|\psi_{1}|^{2} \\
&&-|\nabla_{\mu}\psi_{2}-ie_{2}A_{\mu}\psi_{2}|^{2}-m_{2}^{2}|\psi_{2}|^{2}  \Big),  \nonumber
\end{eqnarray}
where $F_{\mu\nu}=\nabla_{\mu}A_{\nu}-\nabla_{\nu}A_{\mu}$ is the field strength for the $U(1)$ potential $A_{\mu}$. \{$e_{1}$, $m_{1}$\} and \{$e_{2}$, $m_{2}$\} are the charges and masses of the scalar fields $\psi_{1}$ and $\psi_{2}$, respectively. The consistent ansatz for the matter fields can be taken as
\begin{eqnarray}
\psi_{1}&=\psi_{1}(r) \,,
\qquad
\psi_{2}=\psi_{2}(r) \,,
\qquad
A=\phi(r)\text{d}t \,.
\end{eqnarray}
where $\psi_{1},\psi_{2},\phi$ are all functions with respect to r only. In the rest of this paper, we set the AdS radius $L=1$. Then the equations of motion of matter fields on the AdS black brane background are
\begin{equation}\label{eq:psi1}
\psi_{1}''+(\frac{3}{r}+\frac{f'}{f})\psi_{1}'+(\frac{e_{1}^{2}\phi^{2}}{f^{2}}-\frac{m_{1}^{2}}{f})\psi_{1}=0 ,
\end{equation}
\begin{equation}\label{eq:psi2}
\psi_{2}''+(\frac{3}{r}+\frac{f'}{f})\psi_{2}'+(\frac{e_{2}^{2}\phi^{2}}{f^{2}}-\frac{m_{2}^{2}}{f})\psi_{2}=0 ,
\end{equation}
\begin{equation}\label{eq:phi}
\phi''+\frac{3}{r}\phi'-\frac{2(e_{1}^{2}\psi_{1}^{2}+e_{2}^{2}\psi_{2}^{2})}{f}\phi=0 .
\end{equation}

\subsection{Boundary conditions}
In order to solve the above functions, we need to provide suitable boundary conditions both on the horizon $(r=r_{h})$ and on the AdS boundary $(r\rightarrow\infty)$.

The functions $\psi_{1}$, $\psi_{2}$ and $\phi$ can be expanded near the horizon as
\begin{eqnarray}
\psi_{1}&=&\psi_{1h0}+\psi_{1h1}(r-r_{h})+... ,\\
\psi_{2}&=&\psi_{2h0}+\psi_{2h1}(r-r_{h})+...  ,\\
\phi&=&\phi_{h1}(r-r_{h})+\phi_{h2}(r-r_{h})^2+...  ,
\end{eqnarray}
where only $\{ \psi_{1h0}, \psi_{2h0}, \phi_{h1}\}$ are independent.

The expansions near the AdS boundary are
\begin{eqnarray}
\psi_{1}&=&\frac{\psi_{1-}}{r^{\Delta_{1-}}}+\frac{\psi_{1+}}{r^{\Delta_{1+}}} ,\\
\psi_{2}&=&\frac{\psi_{2-}}{r^{\Delta_{2-}}}+\frac{\psi_{2+}}{r^{\Delta_{2+}}} ,\\
\phi&=&\mu-\frac{\rho}{r^{2}} ,\label{phiB}
\end{eqnarray}
with
\begin{equation}\label{ConformalDimension}
\Delta_{i\pm}=2\pm2\sqrt{4+m_{i}^{2}L_{\text{eff}}^{2}}, \qquad
i = 1,2 \, .
\end{equation}

Following the AdS/CFT dictionary, $\mu$ and $\rho$ are chemical potential and charge density of the boundary field theory respectively. In this paper, we take $\psi_{i-}$ as the sources and $\psi_{i+}$ as the vacuum expectation values of the boundary scalar operators $\mathcal{O}_i$. In order to make the $U(1)$ symmetry to be broken spontaneously, we impose $\psi_{1-}=\psi_{2-}=0$ to the solutions labeled by the three independent parameters $\{ \psi_{1h0}, \psi_{2h0}, \phi_{h1}\}$.

\subsection{Scaling symmetry}
From the equations \eqref{eq:psi1}\eqref{eq:psi2}\eqref{eq:phi}, the following scaling symmetries can be found
\begin{eqnarray}
r\rightarrow\lambda r,~
f\rightarrow\lambda^{2}f,~
(t,x,y,z)\rightarrow\lambda^{-1}(t,x,y,z)~;	\\
\psi_{1}\rightarrow\lambda\psi_{1},~
\psi_{2}\rightarrow\lambda\psi_{2},~
\phi\rightarrow\lambda\phi,~
e_{1}\rightarrow \frac{e_{1}}{\lambda},~
e_{2}\rightarrow \frac{e_{2}}{\lambda}~.	\label{eq:scaling2}
\end{eqnarray}

Taking advantage of the first scaling symmetry, we can set $r_{h}=1$ for performing numerics. After solving the coupled differential equations and having the numerical solutions, we can use this scaling symmetry again to recover $r_{h}$ to any value. We use the second scaling symmetry to obtain arbitrary values of electric charge for one of the scalar fields 
. With these two scaling symmetries, the numerical work can be greatly simplified.

\subsection{Free energy}
In order to determine which one is thermodynamically favored, we should calculate the free energy of different solutions. In this paper, we work in the grand canonical ensemble, where the chemical potential $\mu$ is fixed. According to the AdS/CFT dictionary, the  Gibbs free energy $\Omega$ of the boundary thermal state is identified with temperature times the Euclidean on-shell action of the bulk solution. In our case, because we work in the probe limit, contributions from the gravity action are equal for different solutions. Therefore, the differences of the Gibbs free energy between the different solutions at a given temperature only come from the the contribution of matter fields
\begin{equation}
\Omega_{m}=TS_{ME} ~,
\end{equation}
where $S_{ME}$ is the Euclidean on-shell action of matter fields on the black brane background. Because we work in the grand canonical ensemble and choose the scalar operator with dimension $\triangle_{+}$, there is no additional surface term. Thus $\Omega_{m}$ can be calculated by the following expression
\begin{equation}\label{eq:2}
\Omega_{m}=\frac{V_{3}}{16\pi G}(-\frac{1}{2}\mu\rho-\int_{r_{+}}^{\infty}\frac{r^{3}(e_{1}^{2}\psi_{1}^{2}+e_{2}^{2}\psi_{2}^{2})\phi^{2}}{f}\text{d}r) ~,
\end{equation}
where $\mu$, $\rho$ are read from the boundary behavior of the function $\phi(r)$ in \eqref{phiB}, and $V_{3}=\int dxdydz$. In Einstein-Gauss-Bonnet gravity, the effective AdS radius is modified to be $L_{\text{eff}}$. As a result, the volume of the boundary $3-$dimensional space no longer equals $V_3$. We can see from the metric~\eqref{metric} that near the boundary, the metric approaches the following type
\begin{equation}
\text{d}s^{2}=-\frac{r^2}{L_{\text{eff}}^2}\text{d}t^{2}+\frac{L_{\text{eff}}^2}{r^2} \text{d}r^{2}+\frac{r^{2}}{L^{2}}(\text{d}x^{2}+\text{d}y^{2}+\text{d}z^{2}) ,
\end{equation}
where the $t$ coordinate is in accordance with the calculation for the temperature~\eqref{temperature}. Therefore, the boundary spacial coordinates also need to be rescaled as $x=(L/L_{\text{eff}})\bar{x}$ to keep the Lorentz symmetry of the boundary field theory. The boundary volume should be rescaled to be $V_{b3}=\int d\bar{x}d\bar{y}d\bar{z}=(L_{\text{eff}}/L)^3 V_3$. However, we still draw the numerical value of $16 \Pi G \Omega_m/(V_3 \mu^4)$ to analyze the stability of different solutions, because we always compare solutions with the same value of $\alpha$ and $L_{\text{eff}}$.


\subsection{The various solutions}
In this holographic model, there exist solutions dual to the normal phase as well as the condensed phases. The solution dual to the normal phase takes the analytical form
\begin{eqnarray}
\psi_{1}(r)=\psi_{2}(r)=0 \,,
\qquad
\phi(r)=\mu(1-\frac{r_{h}^{2}}{r^{2}})  \,.
\end{eqnarray}
Because there are two s-wave orders in this model, we can obtain three different solutions dual to condensed phases:
\begin{itemize}
  \item \textbf{Solution $S_1$}: $\psi_{1}(r)\neq0, ~\psi_{2}(r)=0$,
  \item \textbf{Solution $S_2$}: $\psi_{1}(r)=0, ~\psi_{2}(r)\neq0$,
  \item \textbf{Solution $S_1+S_2$}: $\psi_{1}(r)\neq0, ~\psi_{2}(r)\neq0$.
\end{itemize}
It has been indicated in Refs.~\cite{Gubser:2008wv,Horowitz:2010gk} that for holographic superconductor models, there exist a discrete family of normalizable solutions that can be labeled by the number of nodes for the condensed fields as functions of $r$. In the three different solutions studied in this work, we only consider the solutions with zero nodes, because other solutions with more nodes are unstable.

In this paper, we aim to find new phase transition behavior of the condensed phase dual to \textbf{Solution $S_{1}+S_{2}$}. According to previous studies~\cite{Nie:2013sda,Nie:2015zia}, \textbf{Solution $S_{1}+S_{2}$} always exist when the free energy curves of \textbf{Solution $S_{1}$} and \textbf{Solution $S_{2}$} have at least one intersection point. Therefore, we need to compare the free energy curves of \textbf{Solution $S_{1}$} and \textbf{Solution $S_{2}$} with different values of \{$e_{1}$,$m_{1}^{2}L_{\text{eff}}^{2}$\} and \{$e_{2}$, $m_{2}^{2}L_{\text{eff}}^{2}$\}. Furthermore, we can see that the equations of motion for $\psi_1$ and $\psi_2$ (\ref{eq:psi1} and \ref{eq:psi2}) have the similar configuration, so we only need to study \textbf{Solution $S_{1}$} with various values of \{$e_{1}$,$m_{1}^{2}L_{\text{eff}}^{2}$\}. The free energy curve of \textbf{Solution $S_{2}$} is same to that of \textbf{Solution $S_{1}$} with the same values of $e_i$ and $m_i$.

\section{The free energy curve of Solution $S_1$}
\label{sec:S1}
The s-wave phase dual to \textbf{Solution $S_{1}$} from Einstein-Gauss Bonnet gravity has already been studied in Refs.~\cite{Gregory:2009fj,Pan:2009xa,Barclay:2010nm,Barclay:2010up,Gregory:2010yr}. But the free energy curves in probe limit have not been analyzed systematically. The previous studies also omit the influence of $e_{1}$ on the condensed behavior of \textbf{Solution $S_{1}$} in probe limit. That is because \textbf{Solution $S_{1}$} with different values of $e_1$ can be get from \textbf{Solution $S_{1}$} with $e_1=1$, taking advantage of the scaling symmetry \eqref{eq:scaling2}. For a holographic system with only one s-wave order, the different values of electric charge will not change the qualitative behavior in probe limit. However, for a system with two s-wave orders, the scaling symmetry can only change the values of two electric charges \{$e_{1}, e_{2}$\} simultaneously, while preserving the ratio of the two. In order to reveal the competition and coexistence between the two s-wave orders, we have to systematically study the influence of both the two parameters \{$e_1,m_1$\} on the free energy curve of \textbf{Solution $S_{1}$}.

The Gauss-Bonnet coefficient $\alpha$ can also change the free energy curve of \textbf{Solution $S_{1}$}. However, when we compare \textbf{Solution $S_{1}$} with \textbf{Solution $S_{2}$} and consider the competition and coexistence between the two orders, the Gauss-Bonnet coefficient should take the same value, while the values of $e_i$ and $m_i$ could be different. Therefore, in the rest of this section, we fix $\alpha$ to five different values, i.e. $\alpha=\{-0.1, 0.0001, 0.1, 0.2, 0.25$\}.At each value of $\alpha$, we plot $\Omega_m$ with respect to temperature for \textbf{Solution $S_{1}$} with different values of \{$e_{1},m_{1}$\}, and study the influence of the two parameters.

It should be noticed that in the probe limit, there is a famous constraint $\alpha\leq 0.09$ on the Gauss-Bonnet coefficient from the boundary causality～\cite{Brigante:2007nu,Brigante:2008gz}. However, some previous studies already revealed some interesting behavior near the Chern-Simons limit~\cite{Gregory:2009fj,Pan:2009xa,Barclay:2010nm,Barclay:2010up,Gregory:2010yr}, which is beyond this causality bound. To make more complete analysis for various values of the Gauss-Bonnet coefficient, and avoid missing interesting results that might be useful in future research, we will also study the cases with $\alpha$ beyond this causality bound. Moreover, it is also interesting to understand the phase structure of asymptotically AdS system from a pure gravity point of view.

\begin{figure}
\begin{center}
\includegraphics[width=4.5cm] {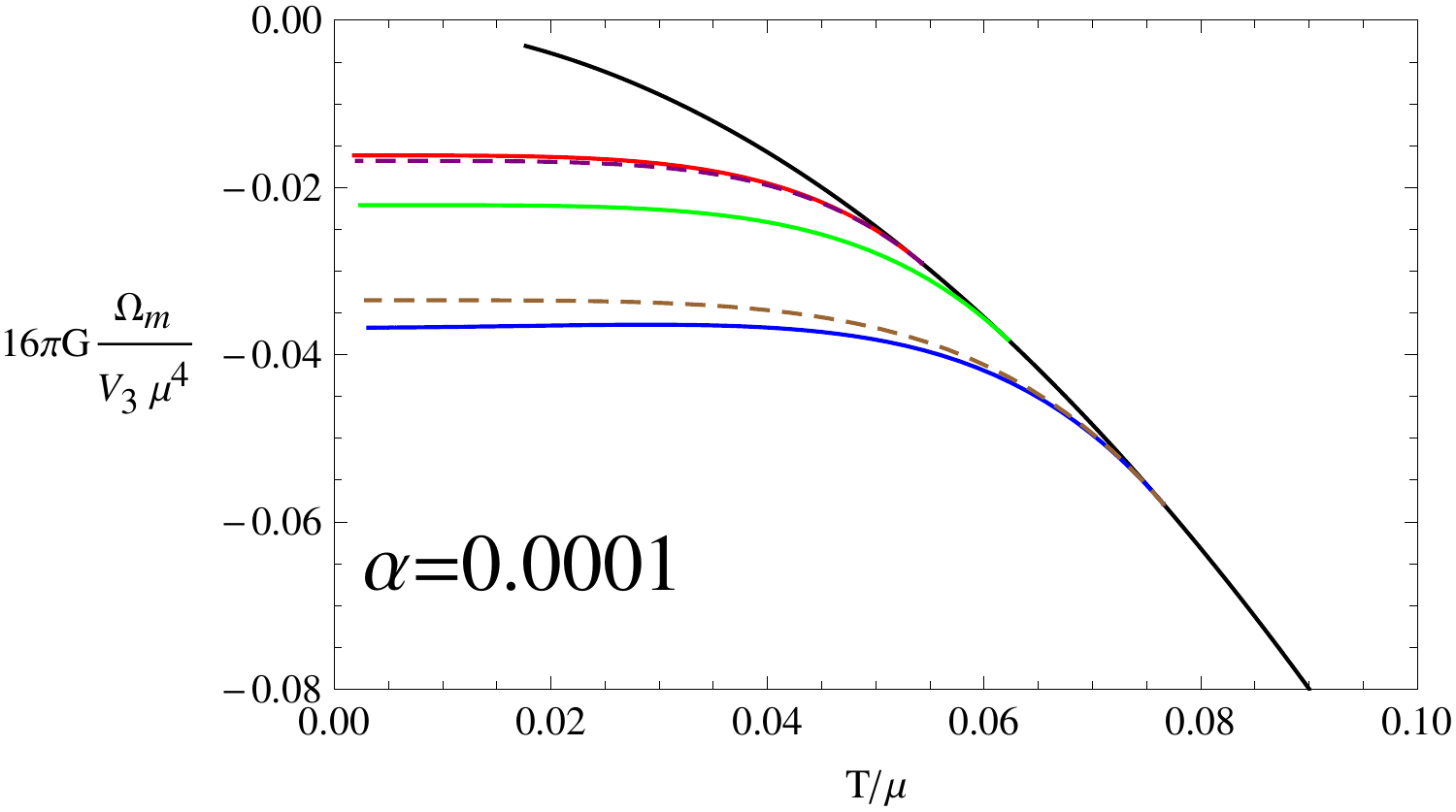}
\end{center}
\caption{\label{e1m1alpha00001}(Color online.)~The Gibbs free energy as a function of temperature with $\alpha=0.0001$. The solid black line corresponds to the normal phase, and the three colored solid lines correspond to \textbf{Solution $S_1$} with fixed value of $e_1=1$ and $m_{1}^{2}L_{\text{eff}}^{2}=-3(blue)$, $-2(green)$, $-1(red)$, respectively. The two colored dashed lines correspond to \textbf{Solution $S_1$} with fixed value of $m_{1}^{2}L_{\text{eff}}^{2}=-2$ and $e_1=0.8721$ (purple), $1.2305$ (brown), respectively. Each dashed line has the same critical temperature with its nearby solid line.}
\end{figure}

We start from the case with $\alpha=0.0001$, and plot the free energy curves for \textbf{Solution $S_{1}$} with various values of \{$e_{1},m_{1}$\} in Figure~\ref{e1m1alpha00001}. In this figure, the solid black line corresponds to the normal phase, and the three colored solid lines correspond to \textbf{Solution $S_1$} with fixed value of $e_1=1$ and $m_{1}^{2}L_{\text{eff}}^{2}=-3(blue)$, $-2(green)$, $-1(red)$, respectively. The two colored dashed lines correspond to \textbf{Solution $S_1$} with fixed value of $m_{1}^{2}L_{\text{eff}}^{2}=-2$ and $e_1=0.8721$ (purple), $1.2305$ (brown), respectively.

From this figure we can see that the three colored solid lines show the influence of the parameter $m_{1}^{2}L_{\text{eff}}^{2}$ on the free energy of \textbf{Solution $S_1$} with fixed values of $\alpha$ and $e_1$. The condensed solution with larger value of $m_{1}^{2}L_{\text{eff}}^{2}$ get higher value of free energy and lower value of critical temperature, which indicates that \textbf{Solution $S_1$} with a lower value of $m_{1}^{2}L_{\text{eff}}^{2}$ is more stable.

We can also combine the two dashed lines with the solid green line into a group. These three lines show the influence of the parameter $e_1$ with fixed values of $\alpha$ and $m_{1}^{2}L_{\text{eff}}^{2}$. We can see that \textbf{Solution $S_1$} with a larger value of $e_1$ get lower value of free energy and higher value of critical temperature, which indicates that a larger value of $e_1$ makes the condensate easier to occur.

From the above analysis, the two parameters $e_1$ and $m_{1}^{2}L_{\text{eff}}^{2}$ have similar influence on the free energy curve of \textbf{Solution $S_1$}. It is obvious that if we only change one of the two parameters, the resulting free energy curves for \textbf{Solution $S_1$} will not intersect with each other, and we can not get \textbf{Solution $S_1+S_2$} dual to the condensed phase with two orders. Therefore, in order to find \textbf{Solution $S_1+S_2$}, we need to make the free energy curves of \textbf{Solution $S_1$} and  \textbf{Solution $S_2$} intersect with each other, and the values of the two parameters $e_2, m_2$ must both differer to the values of $e_1, m_1$.

The value of $e_1$ for the two dashed lines in Figure~\ref{e1m1alpha00001} are chosen to the special values such that its critical temperature is the same to the nearby colored solid line. Therefore, we can also compare the different influence of the two parameters $e_1$ and $m_{1}^{2}L_{\text{eff}}^{2}$ on the free energy curve in this figure. The dashed purple line and the solid red line have the same value of critical temperature, and they are obtained from the solid green line by tuning the value of $e_1$ or $m_{1}^{2}L_{\text{eff}}^{2}$ respectively. The difference between the two lines show that when the critical temperature is decreased by tuning the different parameters to the same value, the resulting free energy curve by tuning $e_1$ is lower than the curve by tuning $m_{1}^{2}L_{\text{eff}}^{2}$, and the difference between the two curves is larger at lower temperature. Along with the relation between the lower two lines (dashed brown line and solid blue line), we can conclude that when the changes of critical temperature are the same, the change of free energy caused by tuning $e_1$ is smaller than that caused by tuning $m_{1}^{2}L_{\text{eff}}^{2}$, and the discrepancy is larger at lower temperature.

\begin{figure}
\includegraphics[width=4.3cm] {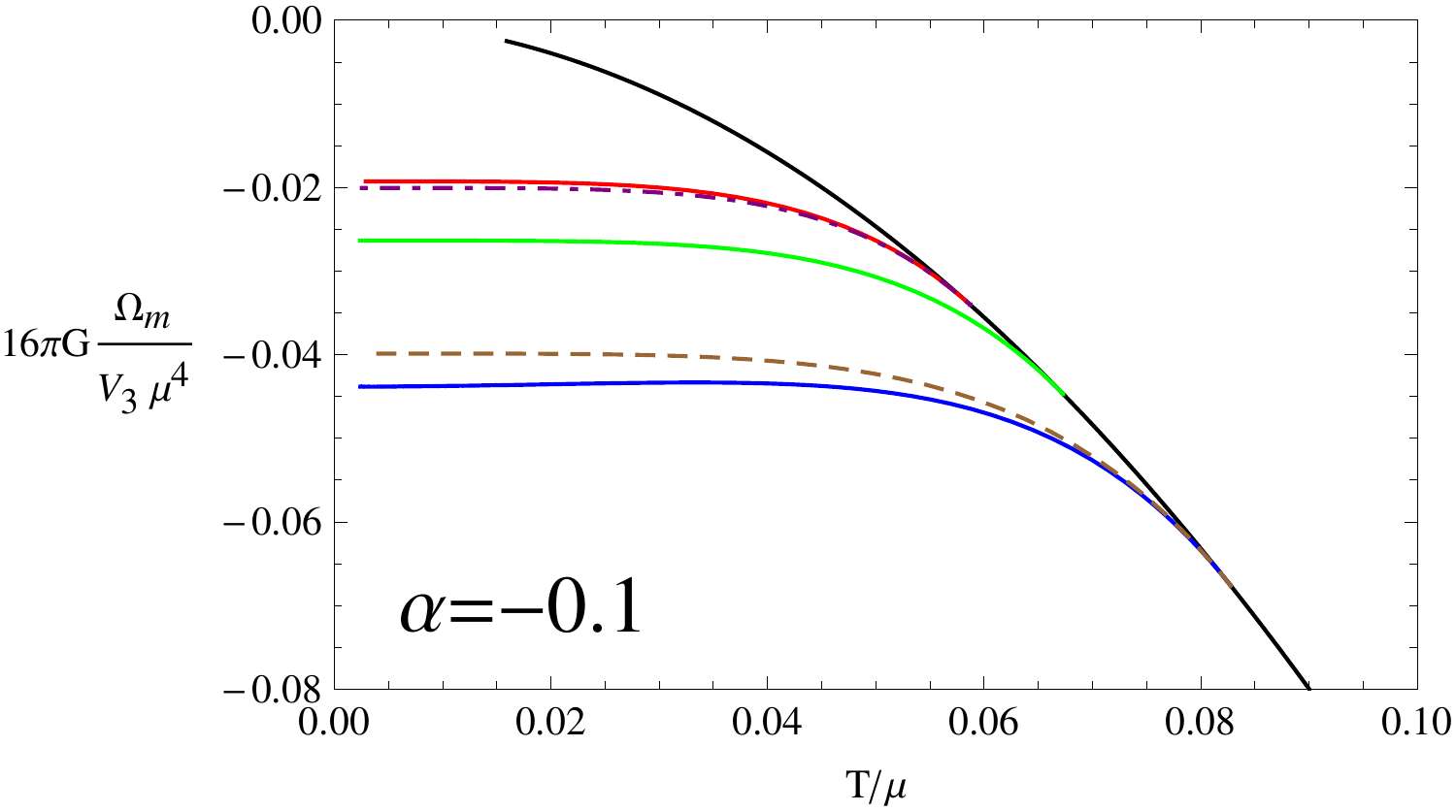}
\includegraphics[width=4.3cm] {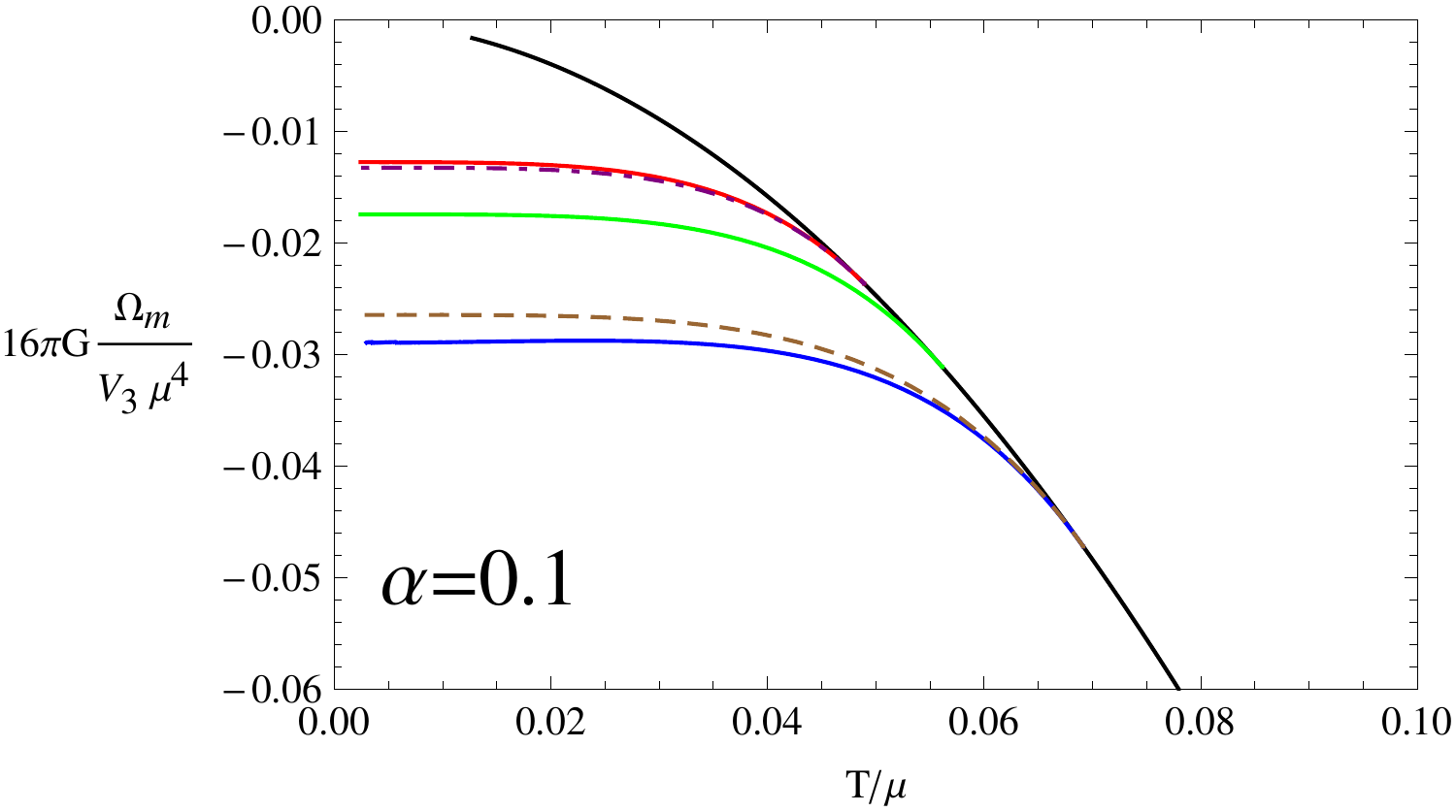}
\includegraphics[width=4.3cm] {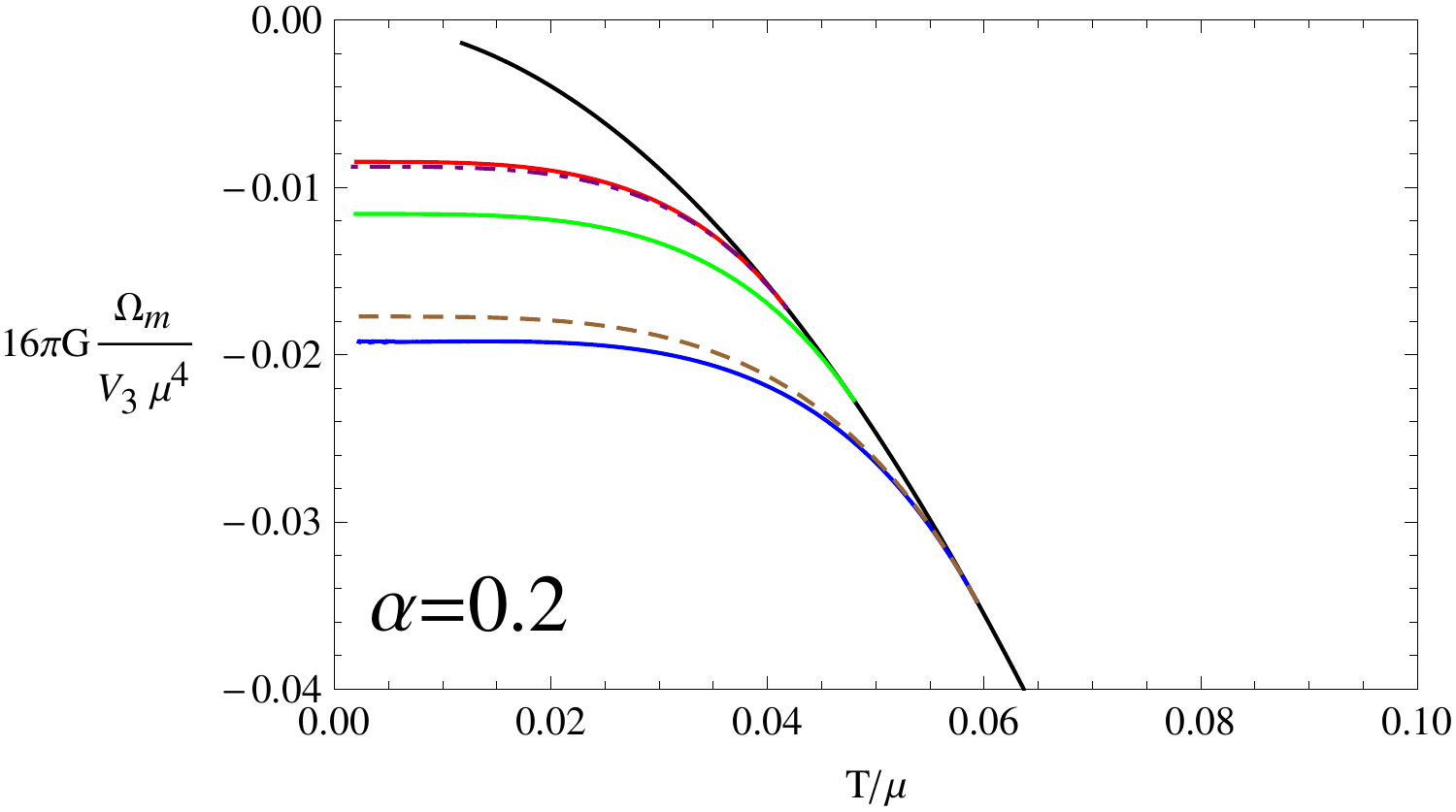}
\includegraphics[width=4.3cm] {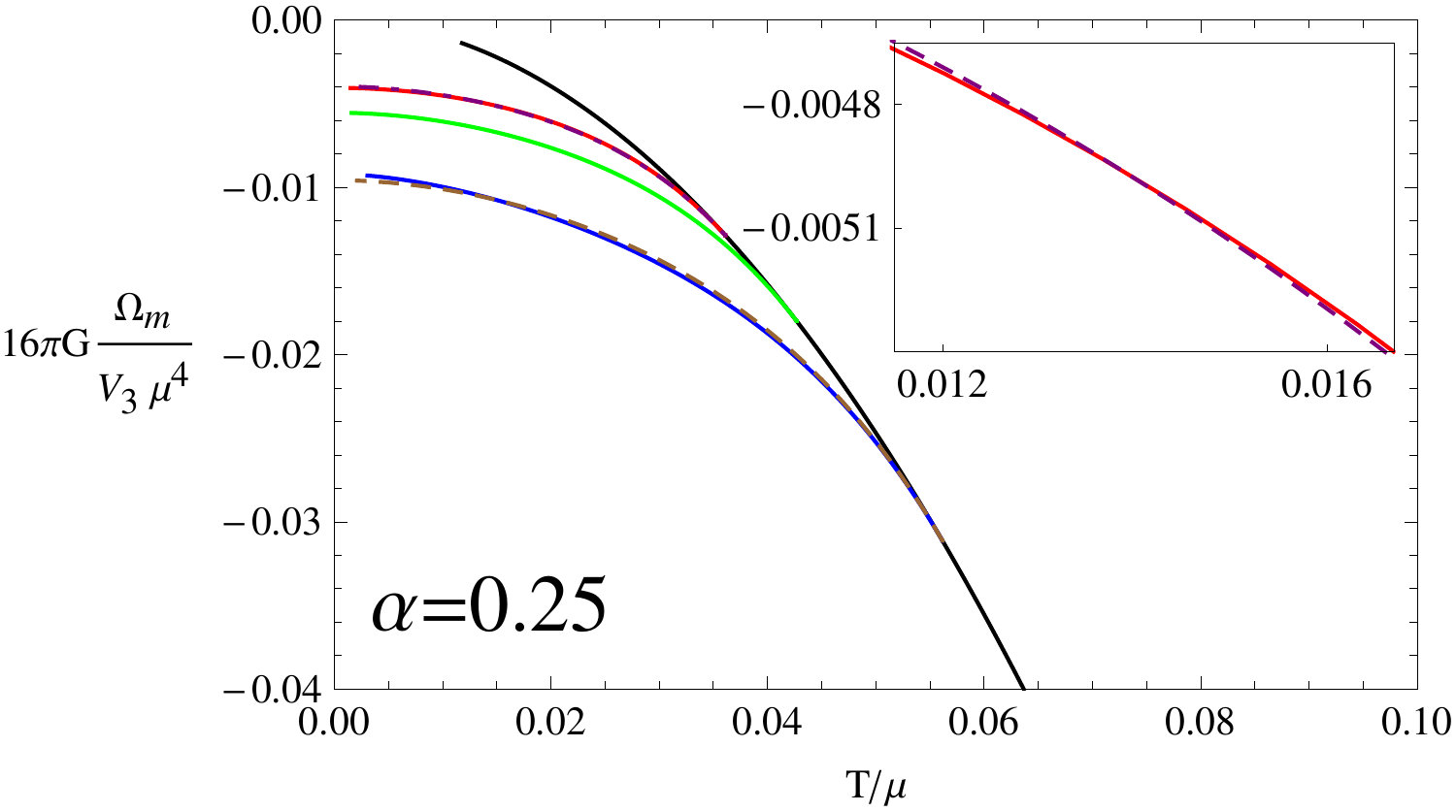}
\caption{\label{4alphas}(Color online.)~The Gibbs free energy $\Omega_{m}$ as a function of temperature with $\alpha=-0.1$, $0.1$, $0.2$, $0.25$ respectively. In each plot, we use the same notations as in Figure~\ref{e1m1alpha00001} for the solid lines, and the two dashed lines correspond to \textbf{Solution $S_{1}$} with $m_{1}^{2}L_{\text{eff}}^{2}=-2$ and two different values of $e_1$, which are listed in table \ref{e1values}. In the fourth plot, we insert an enlarged figure to show details near the intersection point of dashed brown and solid red curves.}
\end{figure}

We also apply the above analysis to the cases with other values of $\alpha$. In figure \ref{4alphas}, we plot the results of $\alpha=-0.1$, $0.1$, $0.2$, $0.25$ respectively. In each of the four plots, we use the same notations as in Figure~\ref{e1m1alpha00001} for the solid lines. The two dashed lines are obtained from the solid green line by changing $e_1$ to appropriate values, such that the dashed lines have the same critical temperature with the solid red and solid blue lines respectively. The detailed values of $e_1$ for these dashed lines are listed in table \ref{e1values}.


\begin{table}
\caption{\label{e1values}Detailed values for $e_1$.}  
\begin{tabular*}{12cm}{lllll}  
\hline  
$\alpha$ & -0.1 \quad & 0.1 \quad & 0.2 \quad & 0.25 \\  
 \hline  
dashed purple  & 0.8723 & 0.8715  & 0.8695 & 0.8456\\
dashed brown  & 1.2298 & 1.2318  & 1.2368 & 1.3152 \\
\hline  
\end{tabular*}  
\end{table}

We can see from figure \ref{4alphas} that, the three plots with $\alpha=-0.1, 0.1, 0.2$ are similar to figure \ref{e1m1alpha00001}. In these plots, the differences between the dashed curves and their nearby solid curves are larger in lower temperature region. As a result, in these cases, we have only found the s+s phase transitions similar to that in the case with $\alpha=0.0001$.

However, it is obvious that the last plot with $\alpha=0.25$ show some differences to the three previous plots. In this plot, each dashed line and its nearby solid line intersect not only at the phase transition critical point, but also at another point with lower temperature. Take the dashed purple line and the solid red line as an example, because the two lines have two intersection points, the free energy value of the dashed purple line is smaller than that of the solid red line in the higher temperature region between the two intersection points, and is larger in the lower temperature region on the left part. Similar behavior can be found for the dashed brown line and solid blue line, where the dashed one has a larger value of free energy at the higher temperature region, and has a smaller value of free energy at the lower temperature region. We can conclude these results as that when the changes of critical temperature are the same, the change of free energy caused by tuning $e_1$ is smaller than that caused by tuning $m_{1}^{2}L_{\text{eff}}^{2}$ in a higher temperature region, and is larger in a lower temperature region.

This special property makes it possible to find interesting reentrant behavior for the s+s phase transitions. The dashed purple line and the solid red line in the last plot of figure \ref{4alphas} has two intersection points. If we further lower the value of $e_1$ for the dashed purple line, the dashed purple line will move upwards. As a result, the two intersection points will move towards each other, and join to be one point where the two lines are tangent. In this process, the solid red line is lower than the dashed purple both in the higher temperature region and in the lower temperature region. Therefore we can expect a reentrant behavior for the s+s phase transition. We will show details for this special behavior of s+s phase in the next section.

\section{The s+s coexistent phases}
\label{sec:s+s}
In the above section, we studied the the free energy curves of the condensed phases with single s-wave order for various values of $e_i$ and $m_i^{2}L_{\text{eff}}^{2}$. These results provide necessary information for searching \textbf{Solution $S_{1}+S_{2}$} which is dual to the s+s coexistent phase. With the information of free energy curves for \textbf{Solution $S_{1}$} and \textbf{Solution $S_{2}$}, we can easily find \textbf{Solution $S_{1}+S_{2}$} near the intersection point of the two free energy curves. In this section, we show the condensate and free energy curves of the two typical phase transition behaviors of s+s coexistent phase for this holographic model. 

\begin{figure}
\includegraphics[width=4.3cm] {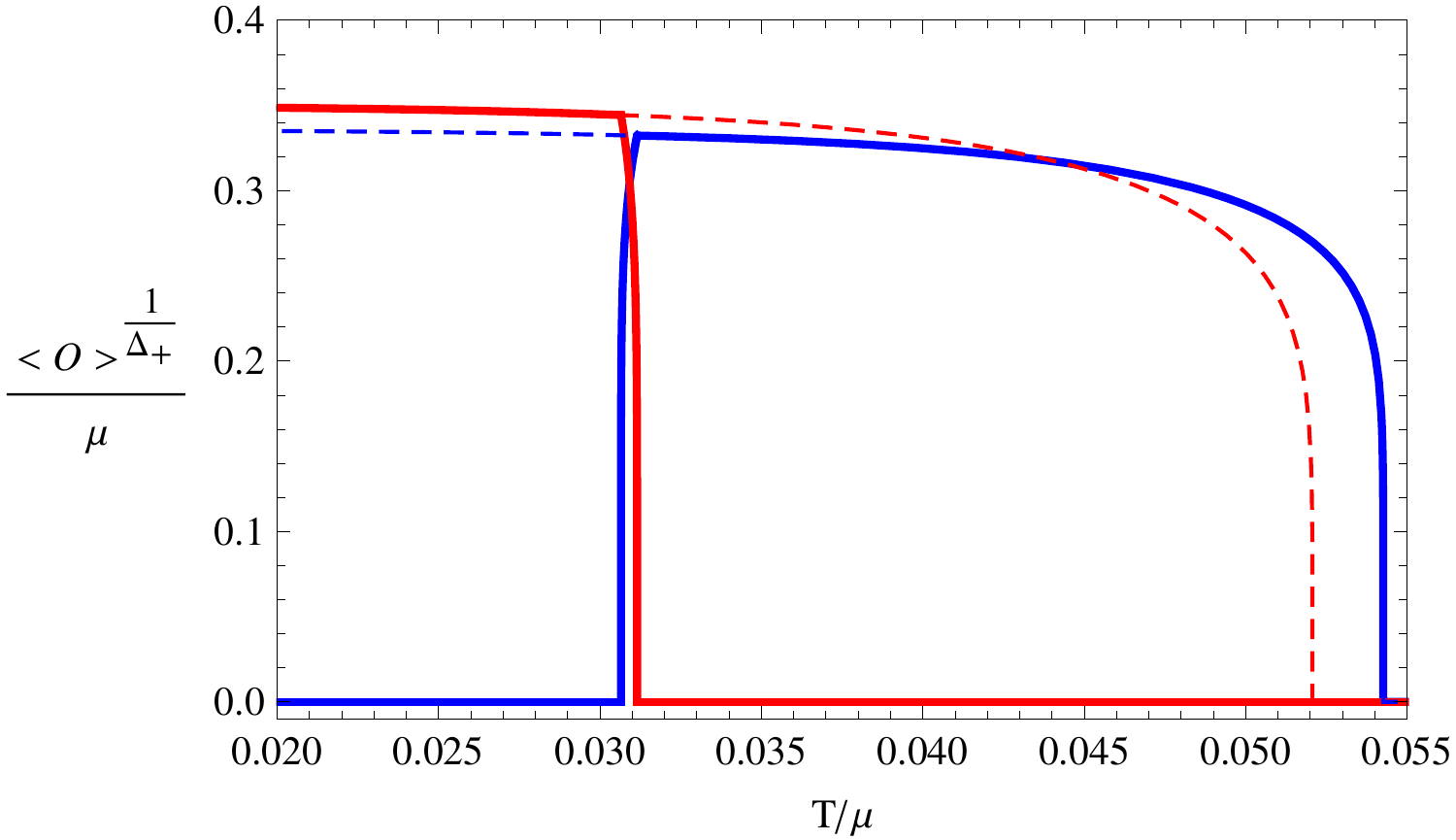}
\includegraphics[width=4.1cm] {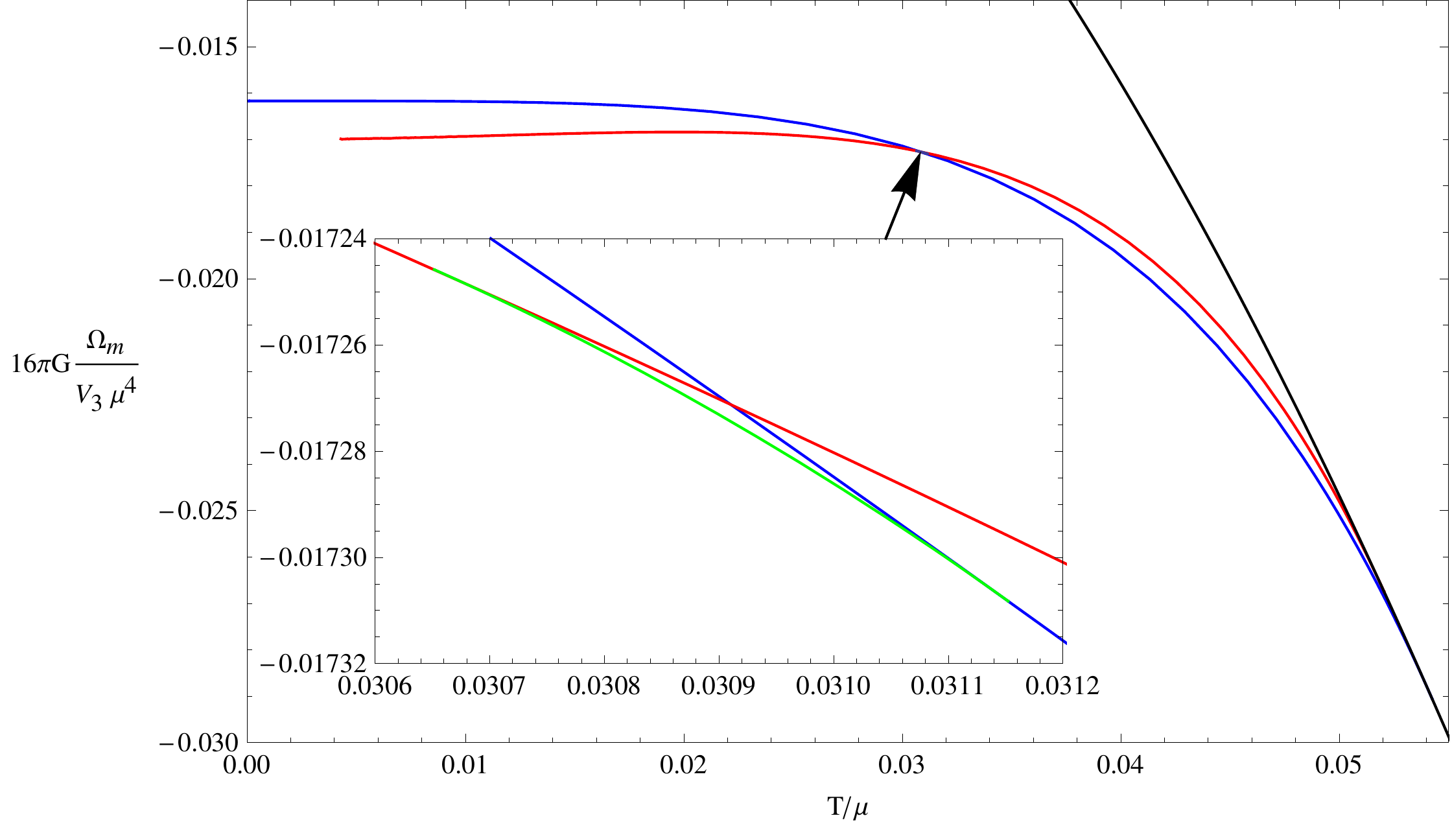}
\caption{\label{xType}(Color online.)~The condensate (left plot) and free energy (right plot) curves with \{$m_{1}^{2}L_{\text{eff}}^{2}=-1$, $e_{1}=1$\}, \{$m_{2}^{2}L_{\text{eff}}^{2}=-3$, $e_{2}=0.68$\} and $\alpha=0.0001$. In the left plot, we use solid blue line to denote the condensate value of the scalar operator dual to $\psi_1$, and use the solid red line to denote the condensate value of that dual to $\psi_2$. Dashed lines denote the condensate value related to $\psi_1$(blue) and $\psi_2$(red) in the unstable sections. In the right plot, the solid blue line corresponds to the free energy curve for \textbf{Solution $S_{1}$}, the solid red line corresponds to the free energy curve for \textbf{Solution $S_{2}$}, and the solid black line is for the normal phase. We insert an enlarged figure to show details near the intersection point of the blue and red lines, where the solid green line denotes the free energy of \textbf{Solution $S_{1}+S_{2}$}.}
\end{figure}

Previous study on holographic systems with two s-wave orders in the probe limit show a typical phase transition behavior for the coexistent phase~\cite{Basu:2010fa}. In our study from Einstein-Gauss-Bonnet gravity in probe limit, this typical behavior of the s+s coexistent phase has also been found. We show the condensate (left plot) and free energy(right plot) curves for this case in figure \ref{xType}, with \{$m_{1}^{2}L_{\text{eff}}^{2}=-1$, $e_{1}=1$\}, \{$m_{2}^{2}L_{\text{eff}}^{2}=-3$, $e_{2}=0.68$\} and $\alpha=0.0001$.

In the left plot of figure \ref{xType}, we use solid blue line to denote the condensate value of the scalar operator dual to $\psi_1$, and use the solid red line to denote the condensate value of that dual to $\psi_2$. The dashed blue and red lines denote the condensate values of the scalar operators dual to $\psi_1$ and $\psi_2$ for unstable sections of \textbf{Solution $S_{1}$} and \textbf{Solution $S_{2}$} respectively. In the right plot, the solid blue line corresponds to the free energy curve for \textbf{Solution $S_{1}$}, the solid red line corresponds to the free energy curve for \textbf{Solution $S_{2}$}, and the solid black line is for the normal phase. We insert an enlarged figure to show details near the intersection point of the blue and red lines, where the solid green line denotes the free energy of \textbf{Solution $S_{1}+S_{2}$}.

From these two plots we can see that \textbf{Solution $S_{1}$} is more stable in the right region with higher temperature, while \textbf{Solution $S_{2}$} is more stable in the left region with lower temperature. \textbf{Solution $S_{1}+S_{2}$} exist in a narrow region near the intersection point of the two free energy curves for \textbf{Solution $S_{1}$} and \textbf{Solution $S_{2}$}, and is the most stable one in this region. Because the condensate curves of the two scalar orders form a shape of letter ``x'', we call this typical phase transition behavior in systems with multiple order parameters as the ``x-type''. In all the values of Gauss-Bonnet coefficient we have studied, we can easily find the ``x-type'' phase transition with suitable values of the parameters \{$m_{1}^{2}L_{\text{eff}}^{2}$, $e_{1}$, $m_{2}^{2}L_{\text{eff}}^{2}$, $e_{2}$\}.

\begin{figure}
\includegraphics[width=4.1cm] {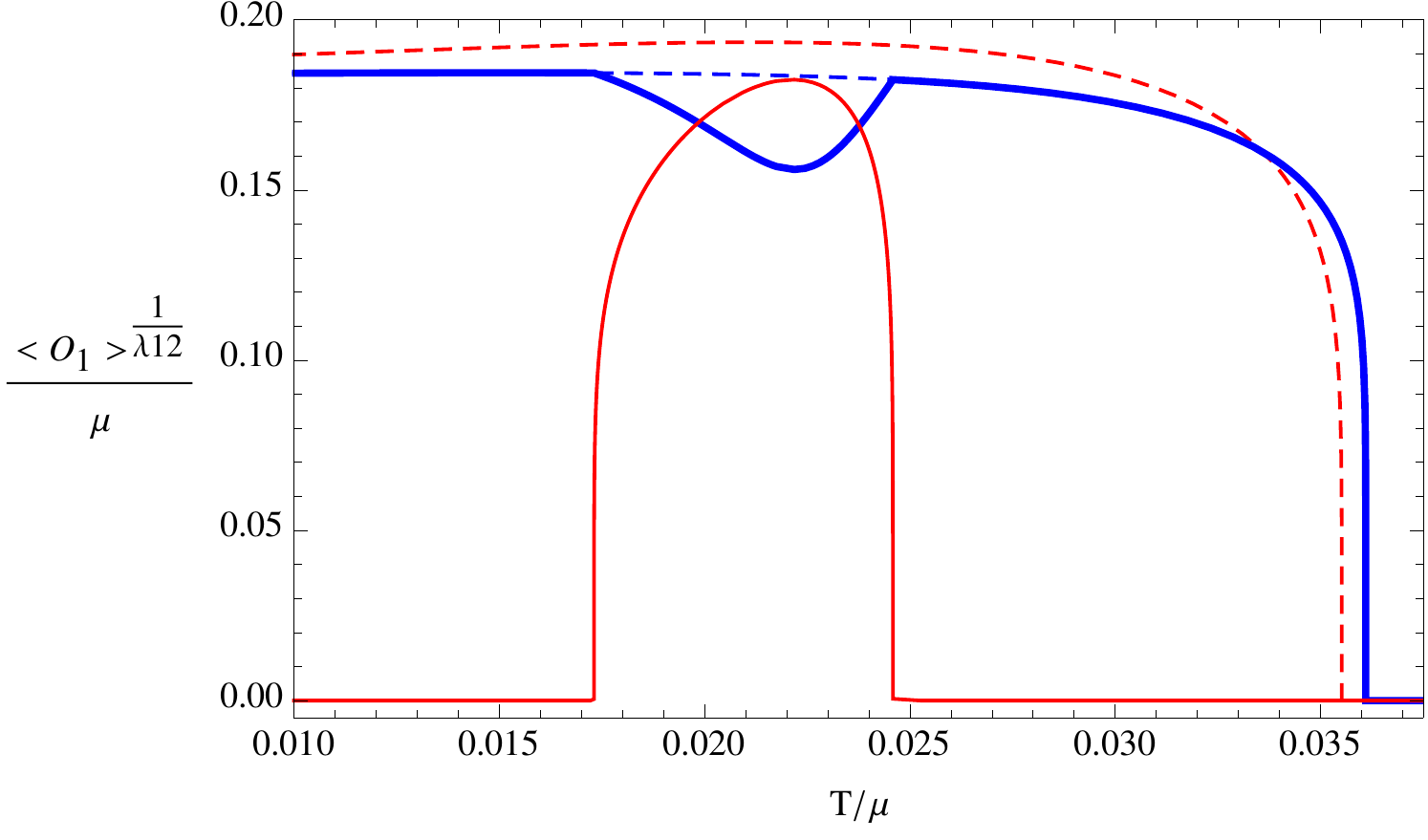}
\includegraphics[width=4.7cm] {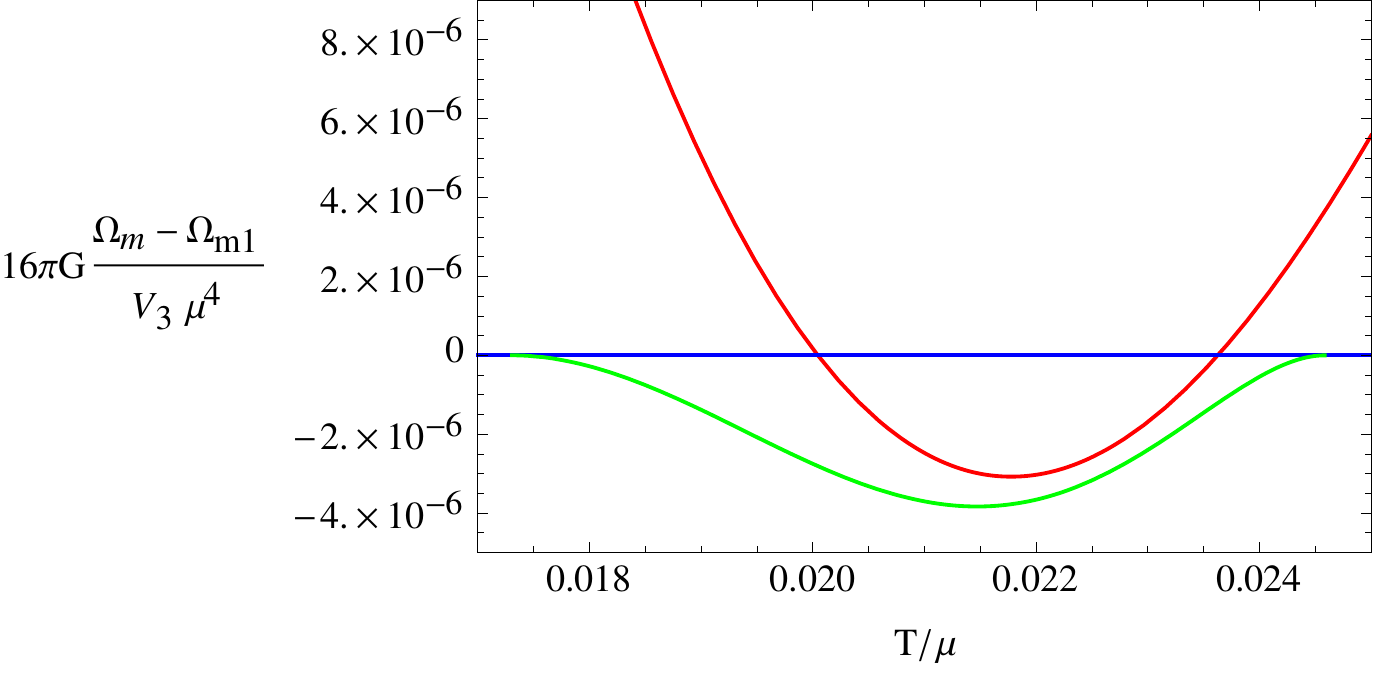}
\caption{\label{nType}(Color online.)~The condensate (left plot) and free energy (right plot) curves with \{$m_{1}^{2}L_{\text{eff}}^{2}=-1 , e_{1}=1$\}, \{$m_{2}^{2}L_{\text{eff}}^{2}=-3 , e_2=0.6325$\} and $\alpha=0.25$.The notations for the curves are the same as those in figure~\ref{xType}.}
\end{figure}

Another kind of phase transition behavior involve with the s+s phase is called the ``n-type'', because one of the condensate curve form a shape of letter ``n''. We draw the condensate (left plot) and free energy (right plot) curves for the ``n-type'' phase transition in figure \ref{nType}, where we use the same notations as in figure \ref{xType}. In the right plot, we show the difference of $\Omega_m$ with respect to that of \textbf{Solution $S_{1}$} ($\Omega_{m1}$), because the absolute values of $\Omega_m$ for different solutions are too close to each other.

From figure \ref{nType} we can see that in this case, \textbf{Solution $S_{1}$} is more stable than \textbf{Solution $S_{2}$} both in the left and right region. Only in the central region, \textbf{Solution $S_{2}$} has lower free energy than \textbf{Solution $S_{1}$}. However, \textbf{Solution $S_{1}+S_2$} exist near this central region, and get the lowest free energy. Therefore \textbf{Solution $S_{2}$} would be always unstable. If we lower the temperature from the right section of \textbf{Solution $S_{1}$}, the system will goes into \textbf{Solution $S_{1}+S_2$} at first, and later goes back to \textbf{Solution $S_{1}$} at a lower temperature. Therefore, this ``n-type'' phase transition is also known as ``reentrance'' in condensed matter physics and has also been found in the holographic s+p system with considering the back reaction on metric~\cite{Nie:2014qma}.

Previous studies on holographic systems with multiple orders in the probe limit only get the ``x-type'' phase transition. Our study show that other interesting phase transition behaviors such as the ``n-type'' one is also possible to be realized from holographic models even in probe limit. We have further confirmed that the conformal dimensions for the two s-wave orders can both be taken to be integers in the ``n-type'' case. These results would be useful in further studies involving non-equilibrium evolution or non homogeneous effects in strongly coupled systems with multiple orders.

\section{Special behavior near the Chern-Simons limit $\alpha=0.25$}\label{sec:alpha0.25}
\begin{figure}
\includegraphics[width=2.7cm] {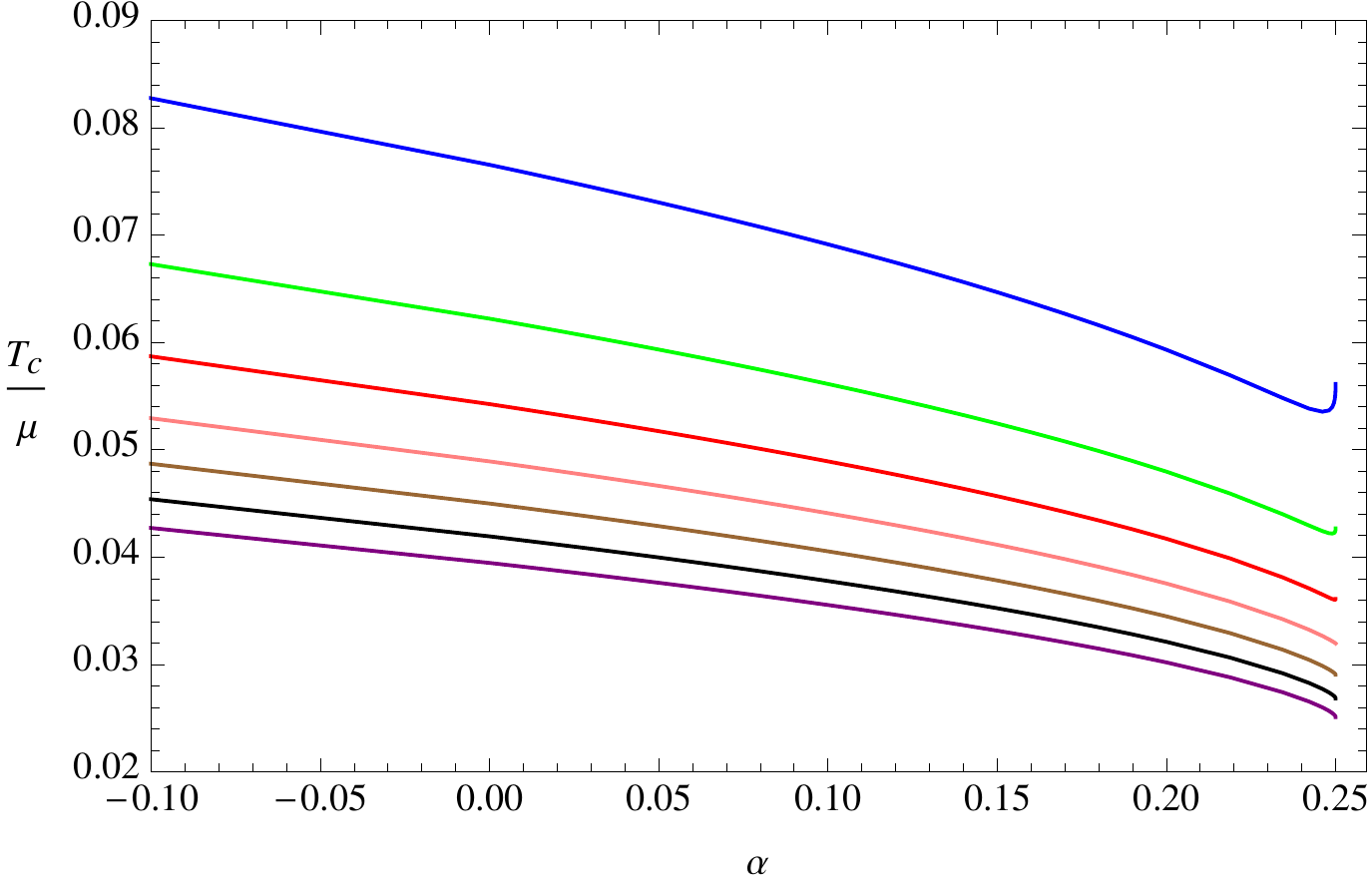}
\includegraphics[width=2.9cm] {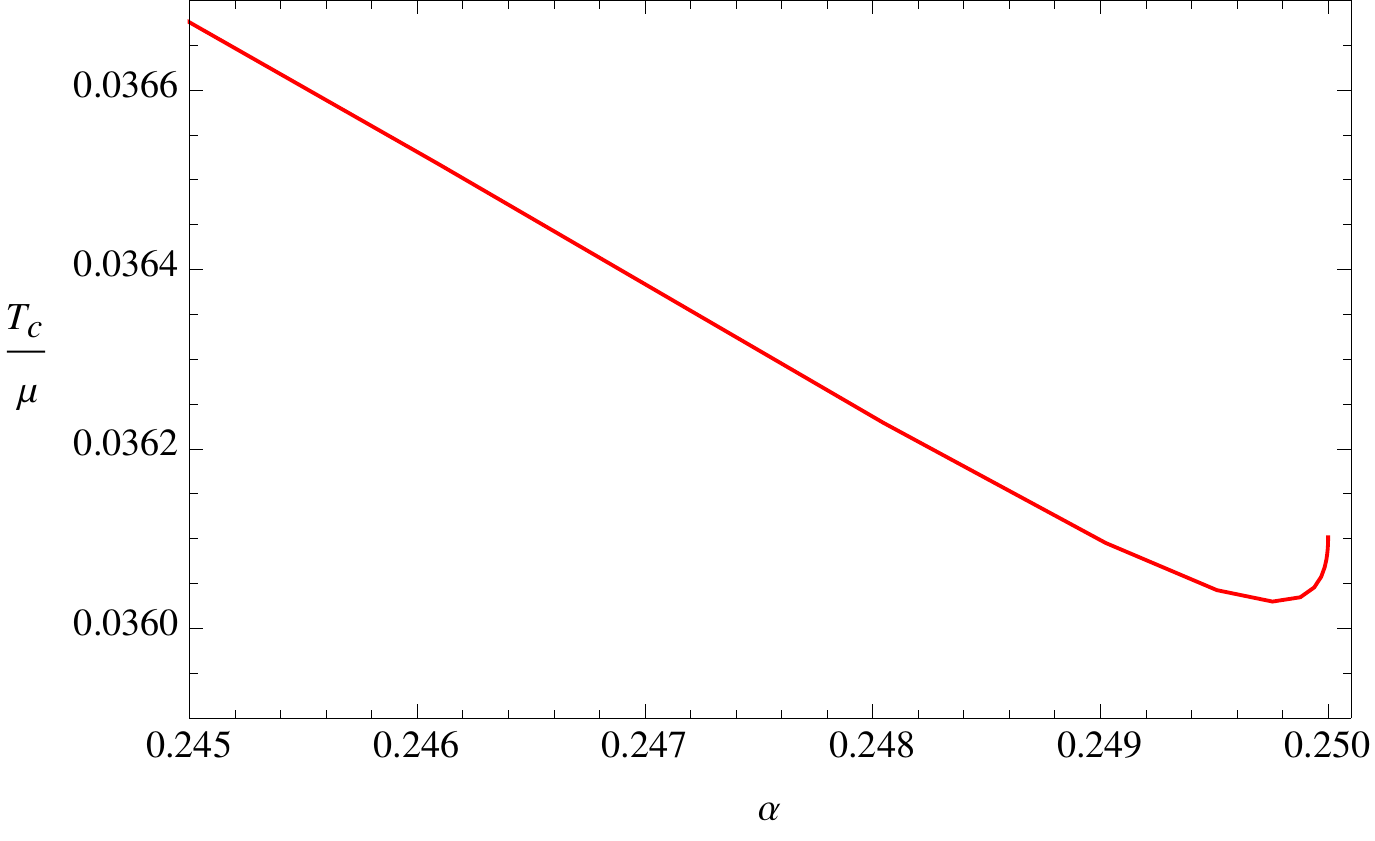}
\includegraphics[width=2.9cm] {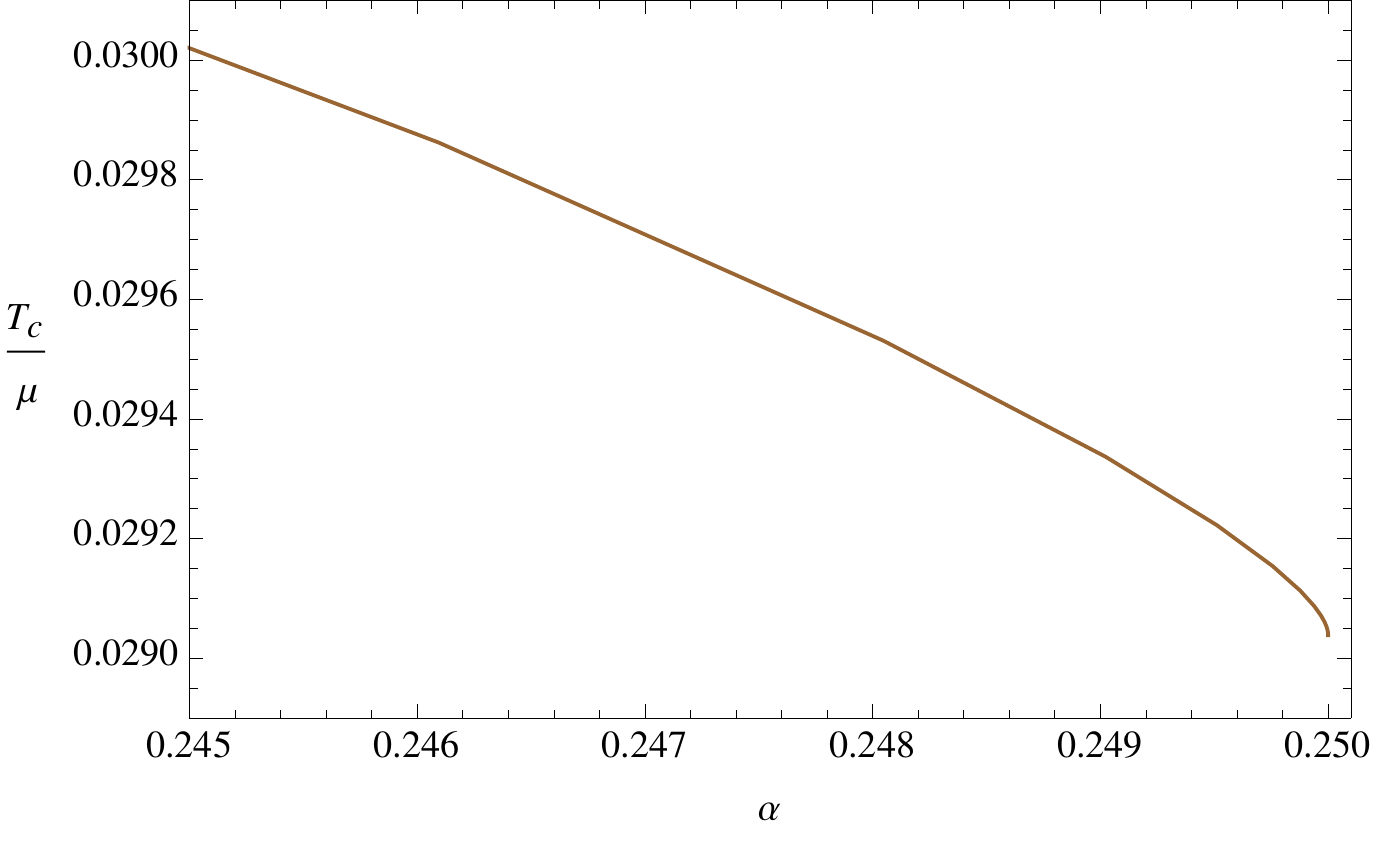}
\caption{\label{Tc-GB}(Color online.)~$T_{c1}-\alpha$ curves for $e_1=1$ and various values of $m_{1}^{2}L_{\text{eff}}^{2}$. The seven colored lines in the first plot from bottom to top correspond to $m_{1}^{2}L_{\text{eff}}^{2}=3(purple)$, $2(black)$, $1(brown)$, $0(pink)$, $-1(red)$, $-2(green)$, $-3(blue)$ respectively. The second and third plots show the tip region near Chern-Simons limit for the red and brown curves in the first plot.}
\end{figure}

It is very strange that the case of $\alpha=0.25$ is quite different to the cases with other values of $\alpha$. We have examined that in a near region below $\alpha=0.25$, we can also found the reentrant behavior for the s+s phase. To better understand this special behavior near the Chern-Simons limit, we recall that in Refs.~\cite{Gregory:2009fj,Pan:2009xa}, the condensate behavior with $\alpha=0.25$ already show some difference. Further studies in Gauss-Bonnet gravity away from probe limit~\cite{Barclay:2010nm,Barclay:2010up,Gregory:2010yr} reveal that the critical temperature of the condensed phase has a nontrivial dependence on the Gauss-Bonnet coefficient near the Chern-Simons limit. This special behavior of $T_c-\alpha$ curve might have some relation to the reentrant behavior of the s+s phase and an explanation of either of the two special behaviors would be heuristic to the understanding of the other one.  Therefore we draw the curves of critical temperature $T_{c1}$ of \textbf{Solution $S_{1}$} as a function of Gauss-Bonnet coefficient with $e_1=1$ and various values of $m_{1}^{2}L_{\text{eff}}^{2}$ in figure \ref{Tc-GB}.

In figure \ref{Tc-GB}, the first plot show how the critical temperatures change when Gauss-Bonnet coefficient $\alpha$ gets different value. We can see from this plot that the critical temperatures generally decrease while Gauss-Bonnet coefficient $\alpha$ is increasing. But when the Gauss-Bonnet coefficient get close to the Chern-Simons limit $\alpha=0.25$, the critical temperatures get a sudden change for most of the seven curves except the pink one with $m_{1}^{2}L_{\text{eff}}^{2}=0$, which is most obvious for the solid blue line. These sudden changes show that the phase transitions near the Chern-Simons limit are affected by some nontrivial effects, which might also lead to the special behavior of free energy curves at $\alpha=0.25$ as well as the reentrant s+s phase transition.

In the second and third plots of figure \ref{Tc-GB}, we show the enlarged version for the red and brown curves near $\alpha=0.25$. We can see from these two plots that the curves with positive and negative values of $m_{1}^{2}L_{\text{eff}}^{2}$ get opposite effects for the sudden change. We also examined the details for all the other five curves and find that the curves with larger absolute value of $m_{1}^{2}L_{\text{eff}}^{2}$ get the sudden change more obviously, and the pink curve with $m_{1}^{2}L_{\text{eff}}^{2}=0$ is almost linear near the Chern-Simons limit. This phenomenon clearly indicates that the special behavior near $\alpha=0.25$ is closely related to some factors in $m_{1}^{2}L_{\text{eff}}^{2}$.

From Eq.~\eqref{Leff2} we can see that $L_{\text{eff}}^2$ depends on $\alpha$, and the derivative $d(L_{\text{eff}}^2)/d\alpha$ diverges while $\alpha \rightarrow 1/4$. It would be reasonable to assume that the sudden change in the critical temperature results from the sudden change of $L_{\text{eff}}^2$ near $\alpha=1/4$. Notice that when we draw the curves in figure \ref{Tc-GB}, we fixed the value of $m_{1}^{2}L_{\text{eff}}^{2}$. Therefore it is the value of $m_1^2$ that changes dramatically with $\alpha$ near the Chern-Simons limit and plays important role in the sudden change of critical temperature.

In the studies of holographic superconductors from Einstein-Gauss-Bonnet gravity, we prefer to fix the value of $m_{1}^{2}L_{\text{eff}}^{2}$ rather than to fix the value of $m_1^2$. This is because that the conformal dimension of the operators in the dual field theory are determined by $m_{1}^{2}L_{\text{eff}}^{2}$ as in Eq~\eqref{ConformalDimension}. Once we fixed the value of $m_{1}^{2}L_{\text{eff}}^{2}$ and study the dependence of critical temperature on Gauss-Bonnet coefficient, the value of $m_1^2$ changes with $\alpha$. From the bulk point of view, the instability that lead to the condensate of scalar field is closely related to dynamics of the scalar field near the horizon, which is determined by both the near horizon geometry and the scalar mass $m_1^2$. When the value of $\alpha$ approaches the Chern-Simons limit, the near horizon geometry changes linearly, while the asymptotic geometry (determined by $L_{\text{eff}}^{2}$) changes dramatically. Therefore the value of $m_1^2$ also changes dramatically and this finally results in a sudden change of critical temperature.

\begin{figure}
\centering
\includegraphics[width=4.3cm] {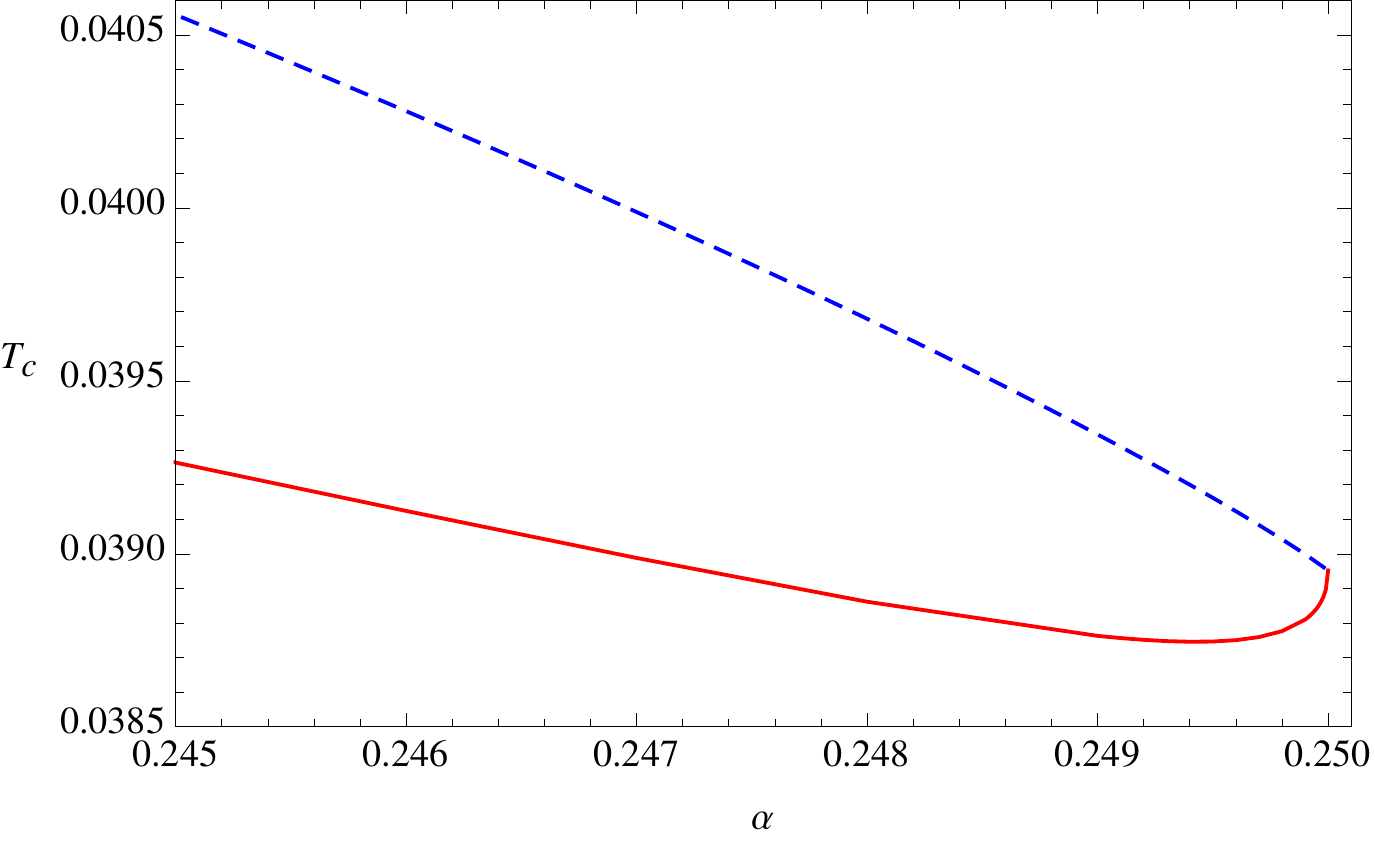}
\caption{\label{Tc-GB-m1}(Color online.)~$T_{c1}-\alpha$ curves with fixed value of $e_1=1$ for $m_1^2=-3$(dashed blue) and $m_{1}^{2}L_{\text{eff}}^{2}=-3/2$(solid red), respectively.}
\end{figure}

To verify the above analysis, we compare the $T_{c1}-\alpha$ curves with fixed values of $m_{1}^{2}L_{\text{eff}}^{2}$ and $m_1^2$ in figure \ref{Tc-GB-m1}. In this figure, the dashed blue line denotes the curve with fixed value of $m_1^2=-3$ and the solid red line denotes that with fixed value of $m_{1}^{2}L_{\text{eff}}^{2}=-3/2$. For both the two curves, the value of $e_1$ are fixed to $e_1=1$. We can see that near the Chern-Simons limit, the curve with fixed $m_1^2$ changes almost linearly, while the curve with fixed $m_{1}^{2}L_{\text{eff}}^{2}$ changes dramatically. This is consistent with our analysis.

The above study addressed the reason of the sudden change of $T_{c1}-\alpha$ curves with fixed value of $m_{1}^{2}L_{\text{eff}}^{2}$ to the sudden change of $L_{\text{eff}}^2$. Although with this special behavior of $T_{c1}-\alpha$ curve, we can not deduce the special behavior of free energy curves away from the critical point and the reentrant phase transition near Chern-Simons limit, our analysis would still be heuristic to the problem. The $T_{c1}-\alpha$ curves only show the influence near the critical region, one might include the analysis to the very low temperature region to get more clues.

\section{Conclusions and discussions}
\label{sec:Conclusions}
In this paper, the holographic superconductor model with two s-wave orders from Einstein-Gauss-Bonnet gravity is studied in the probe limit. The influence of the parameters \{$e_1$, $m_1^2 L_{\text{eff}}^2$\} on the free energy curve of \textbf{Solution $S_{1}$}, which is dual to the condensed phase with only one order, is studied systematically. It is found that by tuning either of the two parameters, the free energy curve along with the critical temperature can be changed. If we want to increase the critical temperature of the condensed phase, we can either increase the value of $e_1$ or decrease the value of $m_1^2 L_{\text{eff}}^2$. In both cases, the condensed solution with increased value of critical temperature will possess lowered value of free energy. However, if only one of the two parameters is changed, the free energy curves of the resulting different solutions will not intersect with each other.

We also compared the different influence of tuning the two different parameters. We fixed the change of critical temperature caused by tuning the two different parameters and compare their influence on the free energy curve. We found that when the Gauss-Bonnet coefficient is far away from the Chern-Simons limit, the influence of tuning the parameter $m_i^2 L_{\text{eff}}^2$ has stronger influence on the change of free energy of the condensed solution, and the difference is larger in the lower temperature region. It is very interesting that when the Gauss-Bonnet coefficient  approaches the Chern-Simons limit, the influence of tuning $m_i^2 L_{\text{eff}}^2$ on the free energy is larger in a higher temperature region and is smaller in the lower temperature region. As a result, the two free energy curves from tuning the different parameter could have two intersection points, and a reentrant ``n-type'' phase transition can be found with appropriate values of the parameters \{$e_1$, $e_2$, $m_1^2 L_{\text{eff}}^2$, $m_2^2 L_{\text{eff}}^2$\}.

With the above analysis on the free energy curves of the condensed solutions with single s-wave order, we find two typical phase transition behaviors of the s+s coexistent phase, and draw their condensate and free energy curves. According to the special shape of the condensate curves, we call the two kinds of phase transitions the ``x-type'' and ``n-type'' respectively. Our study show that we can find not only the typical ``x-type'' phase transitions, but also the reentrant ``n-type'' phase transitions for the s+s phase from holographic models in the probe limit.

The speciality of this holographic model near Chern-Simons limit is not only exhibited in the free energy curves and reentrant phase transitions, but also indicated in the $T_c-\alpha$ curves while $m_i^2 L_{\text{eff}}^2$ is fixed to various values. We explained that the special behavior of $T_c-\alpha$ curves near the Chern-Simons limit is closely related to the sudden change of $L_{\text{eff}}$ when $\alpha$ approaches $0.25$, which greatly changes the asymptotic behavior of the bulk spacetime. In order to keep the conformal dimension of the s-wave operator unchanged, the mass parameter $m_i$ need to be changed accordingly. This finally caused the sudden change in the critical temperature of the condensed solutions with nonzero $m_i$.


There are many further problems need to be studied based on our work. One of these is to seek out the certain relation between intersection of the free energy curves and the existence of coexistent phase. Because we work in probe limit and the conformal dimension of the two s-wave orders can both be taken to be integers in the reentrant case, the numerical work will be greatly simplified in further studies involving time dependence or non homogenous effects based on our work. It would be interesting to realize time dependent solutions across the reentrant s+s phase, and study interesting questions in non-equilibrium systems. One can also build holographic Josephson junction that connect two s-wave part with the same order but exhibit s+s phase near the junction region with a new order. 

\section*{Acknowledgments}
ZYN would like to thank Li-Ming Cao and Qi-Yuan Pan for helpful discussion. This work was supported in part by the National Natural Science Foundation of China under Grant Nos. 11565017, 11247017, 11491240167, 11447131, 11664021, and 11565016, and in part by the Open Project Program of Key Laboratory of Theoretical Physics, Institute of Theoretical Physics, Chinese Academy of Sciences(No. Y5KF161CJ1).

\end{document}